\documentstyle[floats,epsf,eqsecnum,prd,aps,amsfonts]{revtex}
\begin{document}

\draft

%--------------------------------------------------------------------
%  This is the first line to be uncommented for 2 column format
\twocolumn[\hsize\textwidth\columnwidth\hsize\csname
@twocolumnfalse\endcsname
%--------------------------------------------------------------------

\title{Extending the lifetime of 3D black hole computations with a new
hyperbolic system of evolution equations}

\author{Lawrence E. Kidder, Mark A. Scheel, and Saul A. Teukolsky}

\address{Center for Radiophysics and Space Research, Cornell
         University, Ithaca, New York, 14853}

\date{\today}

\maketitle

\begin{abstract}
We present a new many-parameter family of hyperbolic representations
of Einstein's equations, which we obtain by a straightforward
generalization of previously known systems.  We solve the resulting
evolution equations numerically for a Schwarzschild black hole in
three spatial dimensions, and find that the stability of the
simulation is strongly dependent on the form of the equations
(i.e. the choice of parameters of the hyperbolic system), independent
of the numerics. For an appropriate range of parameters we can evolve
a single 3D black hole to $t \simeq 600 M$ -- $1300 M$, and are apparently
limited by constraint-violating solutions of the evolution
equations. We expect that our method should result in comparable times
for evolutions of a binary black hole system.
\end{abstract}


%--------------------------------------------------------------------
% This is the other line to be uncommented for 2 column format
\vskip2pc]
%--------------------------------------------------------------------
%###
%###\narrowtext

\section{Introduction}
\label{sec:Introduction}

A key unsolved problem in general relativity is to provide a detailed
description of the final moments of a binary black hole system as the
two black holes plunge together and merge.  While this problem is
interesting in its own right, the current deployment of LIGO and other
gravitational wave interferometers provides additional incentive for
finding a timely solution: coalescing compact binaries are expected to
be primary sources of gravitational radiation observable by these
instruments.  Comparison of observed gravitational wave forms to
detailed theoretical predictions of binary black hole evolution may
allow one to test general relativity and other theories of
gravitation, to identify black holes in distant galaxies
and to measure their masses and spins.

Although both the initial inspiral of a binary black hole system and
the final ringdown of the resulting Kerr black hole are well described
by perturbation theory, understanding the plunge from the innermost
stable quasicircular orbit through the coalescence will require
numerical solutions of the full Einstein equations in three spatial
dimensions.  Such numerical computations are in
progress\cite{Brandt2000,Alcubierre2001}; however, they are currently
plagued with instabilities that severely limit the duration of the
simulations. Indeed, current 3D Cauchy evolution codes without
built-in symmetries have great difficulty evolving even a single
Schwarzschild black hole for the amount of time that would be required
for a binary orbit.

Many of the stability difficulties affecting black hole computations
are undoubtedly due to the technical details of the numerical solution
scheme; there are many such difficulties to overcome in any large
scale numerical solution of partial differential equations.  However,
there is also evidence that some of the stability problems are due to
properties of the equations themselves, independent of any numerical
approximation. In particular, by rewriting the equations in a
different manner but leaving the numerical method unmodified, one can
significantly affect the stability of the computation
\cite{Alcubierre2001,scheel_etal97b,%
baumgarte99,Alcubierre2000a,Alcubierre2000b,Alcubierre2000}.

Einstein's equations, when written as a Cauchy problem, can be
decomposed into two subsystems of equations: constraint equations that
must be obeyed on each spacelike hypersurface, or time slice, and
evolution equations that describe how quantities propagate from one
hypersurface to the next.  An analogous decomposition occurs in
electromagnetism, which is naturally split into time-independent
(divergence) equations that constrain the fields at a particular time,
and time-dependent (curl) equations that determine their evolution.
For both electromagnetism and gravitation, the system of equations is
overdetermined in the following sense: if the constraint equations are
satisfied at some initial time, then the evolution equations guarantee
that they will be satisfied at subsequent times.  For numerical black
hole computations, one typically solves the constraint equations only
on the initial time slice, and then uses the evolution equations to
advance the solution in time.

However, the decomposition of Einstein's equations into evolution
equations and constraints is not unique. For example, one can add any
combination of constraints to any of the evolution equations to
produce a different decomposition. Indeed, there have been a large
number of new formulations of 3+1 general relativity proposed in
recent years
\cite{baumgarte99,frittelli_reula94,choquet_york95,abrahams_etal95,%
bona_masso95b,mvp96,frittelli_reula96,friedrich96,estabrook_etal97,%
Iriondo1997,anderson_etal98,Bonilla1998,Yoneda1999,Alcubierre1999,%
Frittelli1999,anderson_york99,Friedrich_Rendall2000,Yoneda2000,%
Shinkai2000,Yoneda2001,Hern1999}, many of which have attractive
properties such as symmetric hyperbolicity.

All such formulations must have the same physical solutions since they
describe the same underlying theory.  However, the set of evolution
equations also admits unphysical solutions such as
constraint-violating modes and gauge modes, and these unphysical
solutions will be different for each formulation.  Usually one is not
interested in unphysical solutions, but if such a solution grows
rapidly with time, any small perturbation (say, caused by numerical
errors) that excites this solution will grow and eventually overwhelm
the physical solution.  This is one reason why some formulations of
Einstein's equations may be better suited for numerical evolution than
others.

In order to explore the extent to which different formulations of
Einstein's equations affect the stability of numerical evolutions, we
construct three new formulations of Einstein's equations, following a
method similar to that of \cite{frittelli_reula96}:
\begin{enumerate}
	\item A first-order system obtained directly from the
	      ADM\cite{ADM} system. This system has five undetermined
	      constant parameters that specify constraint terms to be
	      added to the evolution equations. These parameters
	      determine the hyperbolicity of the evolution equations
	      and the values of the characteristic speeds.  We find
	      that constraining the system to have physical
	      characteristic speeds ({\it i.e.,\/} the characteristic
	      fields propagate either along the light cone or normal
	      to the time slice) still leaves two of the five
	      parameters free, and guarantees that the evolution
	      equations are strongly hyperbolic. In this case, the
	      constraint quantities also evolve in a strongly
	      hyperbolic manner with physical characteristic speeds.
	\item A twelve-parameter system obtained by applying a
	      parameterized change of variables to system 1. The
	      additional seven parameters are completely free, and do
	      not affect the hyperbolicity of either the evolution
	      equations or the evolution of the constraint quantities.
	      This system can be reduced to either the Frittelli-Reula
	      formulation\cite{frittelli_reula96} or the
	      Einstein-Christoffel formulation\cite{anderson_york99}
	      with an appropriate choice of the parameters.  The seven
	      additional parameters can be used either to simplify the
	      equations or to improve the numerical behavior of the
	      system.
	\item A two-parameter system that is obtained from system 2 by
	      demanding that the principal part of the equations is
	      equivalent to a scalar wave equation for each of the six
	      components of $g_{ij}$.  This system is particularly
	      simple, is symmetrizable hyperbolic with physical
	      characteristic speeds, and includes the
	      Einstein-Christoffel formulation\cite{anderson_york99}
	      as a special case.
\end{enumerate}

To determine whether modifying the formulation significantly effects
the numerical solution of the evolution equations, we perform
numerical evolutions of single black holes using a new 3D code we have
developed.  We evolve system 3 for simplicity.  We find that by
varying the two parameters in system 3 while keeping the numerical
evolution method fixed, we can vary the run time of the simulation by
more than an order of magnitude.  For a single black hole, our optimum
choice of parameters yields evolutions that run to $t = 600M$ -- $1300M$.
This is long enough that, if this result carries over to
two-black-hole simulations, one could simulate the last few orbits of
a binary system and the final merger.

In section~\ref{sec:Param-Hyperb-Syst} we derive systems 1-3 and
conditions for hyperbolicity. We also derive evolution equations for
the constraint quantities and discuss their hyperbolicity.  In
section~\ref{sec:Numerical-Results} we present numerical evolutions of
system 3 for different choices of parameters, and show that particular
choices yield significant improvements.  In
section~\ref{sec:Discussion} we discuss our results and our plans to
simulate a binary system.

\section{Parameterized Hyperbolic System}
\label{sec:Param-Hyperb-Syst}

\subsection{3+1 ADM}

We begin with the standard 3+1 formulation of \cite{ADM} which is
discussed in detail in \cite{york79}.  Four-dimensional spacetime is
foliated by the level surfaces $\Sigma_t$ of a function $t(x^\mu)$.
Let $n^\mu$ be the unit normal vector to the hypersurfaces $\Sigma_t$.
Then the spacetime metric ${{}^{(4)}g}_{\mu \nu}$ induces the spatial
three-metric $g_{\mu \nu}$ on each $\Sigma_t$ given by
\begin{equation}
g_{\mu \nu} =  {{}^{(4)}g}_{\mu \nu} + n_\mu n_\nu .
\end{equation}
The timelike vector $t^\mu$ is defined such that 
$t^\mu t_{;\mu} = 1$, where $t_{;\mu}$ is the
covariant derivative of $t$ with respect to the spacetime metric.
The lapse function $N$ and shift vector $\beta^\mu$ are defined by
\begin{eqnarray}
N \equiv && -t^\mu n_\mu , \\
\beta_\mu \equiv && g_{\mu \nu} t^\nu .
\end{eqnarray}
If we adopt a coordinate system $\{t,x^i\}$ adapted to the spatial
hypersurfaces, the line element is given in the usual 3+1 form
\begin{equation}
ds^2 = -N^2 dt^2 + g_{ij}(dx^i + \beta^i dt)(dx^j + \beta^j dt).
\end{equation}
The extrinsic curvature $K_{i j}$ of the spatial surfaces is given by
\begin{equation}
\label{eq:excurvdef}
K_{i j} = - \case{1}{2} \pounds_n g_{i j},
\end{equation}
where $\pounds$ denotes a Lie derivative.

Einstein's equations are given in covariant form by
\begin{equation}
{{}^{(4)}\!R}_{\mu \nu} - \case{1}{2} {{}^{(4)}g}_{\mu \nu} {{}^{(4)}\!R} = 
8\pi T_{\mu \nu},
\end{equation}
where ${{}^{(4)}\!R}_{\mu \nu}$ and ${{}^{(4)}\!R}$ are the Ricci tensor and
Ricci scalar associated with the spacetime metric, and $T_{\mu \nu}$
is the stress-energy tensor.  In the 3+1 decomposition Einstein's
equations are decomposed into the Hamiltonian constraint
\begin{equation}
{\cal C} \equiv \case{1}{2} \left( R - K_{a b} K^{a b} + K^2 \right)
- 8\pi\rho = 0,
\end{equation}
the momentum constraints
\begin{equation}
{\cal C}_i \equiv \nabla_a K_i^{~ a} - \nabla_i K - 8\pi J_i = 0,
\end{equation}
and the evolution equations
\begin{eqnarray}
\widehat{\partial_0} K_{i j} = && - \nabla_i \nabla_j N + N R_{i j}
- 2 N K_{i a} K_j^{~ a} + N K K_{i j} \nonumber \\ && \mbox{}
- 8\pi N S_{i j} - 4\pi N g_{i j}(\rho - S),
\end{eqnarray}
where $K = g^{a b} K_{a b}$, and $\nabla_i$, $R_{i j}$, and $R$ are
the covariant derivative, Ricci tensor, and Ricci scalar associated
with the spatial three-metric.  The symbol $\widehat{\partial_0}$ is
the time derivative operator normal to the spatial foliation, defined
by
\begin{equation}
\widehat{\partial_0} \equiv \partial_t - \pounds_\beta.
\end{equation}
The matter terms are defined as
\begin{mathletters}
\begin{eqnarray}
\rho   \equiv && n^\mu n^\nu T_{\mu \nu},\\
J_{i}  \equiv && -n^\mu g^{~\nu}_i T_{\mu \nu},\\
S_{ij} \equiv && g^{~\nu}_i g^{~\mu}_j T_{\mu \nu},
\end{eqnarray}
\end{mathletters}
and $S = g^{a b} S_{a b}$.
The definition~(\ref{eq:excurvdef}) of the
extrinsic curvature yields the following evolution equation for the
spatial metric
\begin{equation}
\label{eq:dgdt}
\widehat{\partial_0} g_{i j} = -2 N K_{i j}.
\end{equation}
Note that the spatial metric and its inverse are used to lower and
raise the indices of all spatial tensors.

\subsection{First-order form}
In order to cast the evolution equations in first-order form,
we must eliminate the second derivatives of the spatial metric.
We define a new variable (symmetric on its last two indices)
\begin{equation}
\label{eq:definitiond}
d_{k i j} \equiv \partial_k g_{i j},
\end{equation}
and its traces $d_k \equiv g^{a b} d_{k a b}$ and $b_k \equiv g^{a b}
d_{a b k}$.
An evolution equation for $d_{k i j}$ is obtained by taking a
spatial derivative of~(\ref{eq:dgdt}) and using the fact that
$\partial_k$ and $\widehat{\partial_0}$ commute.  This yields 
\begin{equation}
\label{eq:dddt}
\widehat{\partial_0} d_{k i j} = -2 N \partial_k K_{i j}
- 2 K_{i j} \partial_k N,
\end{equation}
where the Lie derivative of $d_{k i j}$ is
\begin{eqnarray}
\pounds_{\beta}d_{k i j} &=& \beta^a \partial_a d_{k i j} +
d_{a i j} \partial_k \beta^a + 2 d_{k a (i} \partial_{j)} \beta^a 
 \nonumber \\ && \mbox{} + 2 g_{a (i} \partial_{j)} \partial_k \beta^a .
\end{eqnarray}

Since we have introduced a new variable that we will evolve
independently of the metric, we have an additional constraint,
\begin{equation}
\label{eq:tic}
{\cal C}_{k i j} \equiv d_{k i j} - \partial_k g_{i j} = 0,
\end{equation}
which must be satisfied in order for a solution of the first-order
evolution equations to be a solution of Einstein's equations.
Note that the spatial derivatives
of $d_{k i j}$ must satisfy the constraint
\begin{equation}
\label{eq:fic}
{\cal C}_{k l i j}  \equiv  \partial_{[k} d_{l] i j} = 0,
\end{equation}
because second derivatives of the metric commute.  Therefore we make the
following substitution when we encounter second derivatives of the
metric:
\begin{equation}
\partial_k \partial_l g_{i j} = \partial_{(k} d_{l) i j}.
\end{equation}
In terms of these new variables, the affine connection, Ricci tensor,
and Ricci scalar are given by
\begin{equation}
\Gamma_{k i j} = d_{(i j) k} - \case{1}{2} d_{k i j},
\end{equation}
\begin{eqnarray}
R_{i j} = && \case{1}{2} g^{a b} \left( \partial_{(i} d_{a b j)} 
+ \partial_a d_{(i j) b} - \partial_a d_{b i j} 
- \partial_{(i} d_{j) a b} \right)
\nonumber \\ && \mbox{}
+ \case{1}{2} b^a d_{a i j} - \case{1}{4} d^a d_{a i j} - b^a d_{(i j) a} 
- \case{1}{2} d_{a j}^{~ ~ b} d_{b i}^{~ ~ a} \nonumber \\ && \mbox{}
+ \case{1}{2} d^a d_{(i j) a} + \case{1}{4} d_i^{~ a b} d_{j a b}
+ \case{1}{2} d_{~ ~ i}^{a b} d_{a b j},
\end{eqnarray}
\begin{eqnarray}
R = && g^{a b} g^{c d} \left( \partial_d d_{a b c} - \partial_a d_{b c d}
\right) + b^a d_a - b_a b^a - \case{1}{4} d_a d^a  \nonumber \\ && \mbox{}
- \case{1}{2} d_{a b c} d^{c a b} + \case{3}{4} d_{a b c} d^{a b c}.
\end{eqnarray}
The constraint equations are given by
\begin{eqnarray}
\label{eq:hc}
{\cal C} &=& \case{1}{2} g^{a b} g^{c d} \left( \partial_d d_{a b c} - 
\partial_a d_{b c d} \right) + \case{1}{2} b^a d_a - \case{1}{2} b_a b^a 
\nonumber \\ && \mbox{}- \case{1}{8} d_a d^a 
- \case{1}{4} d_{a b c} d^{c a b} + \case{3}{8} d_{a b c} d^{a b c}
\nonumber \\ && \mbox{} - \case{1}{2} K_{a b} K^{a b} + \case{1}{2} K^2  
- 8\pi\rho,
\end{eqnarray}
\begin{eqnarray}
\label{eq:mc}
{\cal C}_i &=& g^{a b} \left( \partial_a K_{i b} - \partial_i K_{a b} \right)
+ \case{1}{2} K^{a b} d_{i a b} \nonumber \\ && \mbox{}
+ \case{1}{2} K_{i a} d^a - K_{i a} b^a - 8\pi J_i.
\end{eqnarray}
Finally, the evolution equation for the extrinsic curvature becomes
\begin{eqnarray}
\label{eq:dkdt}
\widehat{\partial_0} K_{i j} = && 
N \left[ \case{1}{2} g^{a b} \left( \partial_{(i} d_{a b j)} 
+ \partial_a d_{(i j) b} - \partial_a d_{b i j} 
- \partial_{(i} d_{j) a b} \right) \right. \nonumber \\ && \mbox{}
+ \case{1}{2} b^a d_{a i j} - \case{1}{4} d^a d_{a i j} - b^a d_{(i j) a} 
- \case{1}{2} d_{a j}^{~ ~ b} d_{b i}^{~ ~ a} \nonumber \\ && \mbox{}
+ \case{1}{2} d^a d_{(i j) a} + \case{1}{4} d_i^{~ a b} d_{j a b}
+ \case{1}{2} d_{~ ~ i}^{a b} d_{a b j} - 2 K_{i a} K_j^{~ a} 
\nonumber \\ && \mbox{} \left. + K K_{i j} \right]
- \partial_i \partial_j N - \case{1}{2} d^a_{~ i j} \partial_a N
+ d_{(i j)}^{~~ ~~ a} \partial_a N  \nonumber \\ && \mbox{}
- 8 \pi N S_{i j} - 4\pi N g_{i j}(\rho - S).
\end{eqnarray}

The hyperbolicity of the system of evolution equations can be determined by
examining its principal part.
Consider a system of the form 
\begin{equation}
\label{eq:firstorderform}
\hat{\partial}_0 u + A^i \partial_i u = F,
\end{equation}
where $u$ is a column vector of the fundamental variables, and $A^i$
and $F$ are matrices that can depend on $u$, but not on derivatives of
$u$.  One defines, for a particular unit one-form $\xi_i$, the
characteristic matrix $C$ in the direction normal to $\xi_i$:
\begin{equation}
\label{eq:characteristicmatrix}
C \equiv A^i \xi_i.
\end{equation}
The characteristic speeds in the direction $\xi_i$ are the eigenvalues
of $C$.  If all characteristic speeds are real, then the system is
said to be weakly hyperbolic. If in addition, $C$ has a complete set
of eigenvectors, and the matrix of these eigenvectors and its inverse
are uniformly bounded functions of $\xi_i$, the spacetime
coordinates, and the solution, then the system is said to be strongly
hyperbolic.  If the matrices $A^i$ are symmetric, the system is said
to be symmetric hyperbolic. If the matrices $A^i$ can be brought into
symmetric form by multiplying by a positive-definite matrix called a
symmetrizer, the system is said to be symmetrizable hyperbolic.
Symmetric, symmetrizable, and strongly hyperbolic systems admit a
well-posed Cauchy problem; weakly hyperbolic systems do
not\cite{Kreiss1989}.  

For the systems described in this paper, we explicitly construct a
complete set of eigenvectors that depend upon $\xi_i$, the
metric, and its inverse. Provided that the matrix norms of the metric
and its inverse remain bounded, then the norms of the matrix of
eigenvectors and its inverse are bounded, so the system is
strongly hyperbolic\cite{Stewart1998}.

Using the method outlined in Appendix~\ref{sec:charsystem}, we find
that the ADM equations written in first-order form are only weakly
hyperbolic, as the characteristic matrix of the system has eigenvalues
$\{0,\pm 1\}$, but does not have a complete set of eigenvectors.
Fortunately, the hyperbolicity of the equations can be changed by
``densitizing'' the lapse and adding constraints to the evolution
equations.

\subsection{Densitization of the lapse}
We densitize the lapse by defining
\begin{equation}
Q \equiv \log \left( N g^{-\sigma} \right),
\end{equation}
where $g$ is the determinant of the three-metric, and $\sigma$ is the
densitization parameter, which is an arbitrary constant.  The lapse
density $Q$ and the shift vector $\beta^i$ will be considered as
arbitrary gauge functions {\it independent} of the dynamical fields.
With this definition we have
\begin{eqnarray}
\partial_i N = && N \left( \partial_i Q + \sigma d_i \right), \\ 
\partial_i \partial_j N = && N [ \partial_i \partial_j Q 
+ (\partial_i Q) (\partial_j Q) + 2 \sigma d_{(i} \partial_{j)} Q
\nonumber \\ && \mbox{}
+ \sigma g^{a b} \partial_{(i} d_{j) a b} 
- \sigma d_{i a b} d_j^{~ a b} + \sigma^2 d_i d_j ].
\end{eqnarray}
Substituting the above expressions into the evolution equations, and
examining the hyperbolicity of the modified evolution equations, we
find that densitizing the lapse is not sufficient to make the
evolution system strongly hyperbolic.  In order for the system to
remain even weakly hyperbolic the densitization parameter must satisfy
$\sigma \geq 0$, as the eigenvalues of the characteristic matrix are
now $\{0,\pm 1,\pm \sqrt{2\sigma}\}$.  In the next section, we will
find that densitizing the lapse is a necessary condition for strong
hyperbolicity, and that if we demand physical characteristic speeds we
must choose $\sigma = \case{1}{2}$.

\subsection{Addition of constraints: System 1}
\label{sec:System1}
By adding terms proportional to the constraints, we can modify the
evolution equations for $K_{i j}$ and $d_{k i j}$ without affecting
the physical solution.  We modify the evolution
equations~(\ref{eq:dddt}) and~(\ref{eq:dkdt}) by
\begin{eqnarray}
\widehat{\partial_0} K_{i j} = && \left( \ldots \right) 
+ \gamma N g_{i j} {\cal C} + \zeta N g^{a b} {\cal C}_{a (i j) b}, \\ 
\widehat{\partial_0} d_{k i j} = && \left( \ldots \right) 
+ \eta N g_{k (i} {\cal C}_{j)} + \chi N g_{i j} {\cal C}_k,
\end{eqnarray}
where $\left( \ldots \right)$ represents the right-hand side of
either equation~(\ref{eq:dddt}) or~(\ref{eq:dkdt}), and the constraint
parameters $\{ \gamma, \zeta, \eta, \chi \}$ are arbitrary constants.
The evolution equations are now given by
\begin{eqnarray}
\widehat{\partial_0} g_{i j} \simeq && 0, \\
\widehat{\partial_0} K_{i j} \simeq && 
- \case{1}{2} N g^{a b} \left[ \partial_a d_{b i j}
- (1 + \zeta) \partial_a d_{(i j) b} 
\right. \nonumber \\ && \mbox{} 
- (1 - \zeta) \partial_{(i} d_{a b j)} 
+ (1 + 2 \sigma)  \partial_{(i} d_{j) a b} 
\nonumber \\ && \mbox{} \left.
- \gamma g_{i j} g^{c d} \partial_a d_{c d b}
+ \gamma g_{i j} g^{c d} \partial_a d_{b c d} \right], \\ 
\widehat{\partial_0} d_{k i j} \simeq && 
- 2 N \partial_k K_{i j}
+ N g^{a b} \left( \eta g_{k (i} \partial_a K_{b j)}
+ \chi g_{i j} \partial_a K_{b k}
\right. \nonumber \\ && \mbox{} \left.
- \eta g_{k (i} \partial_{j)} K_{a b}
- \chi  g_{i j} \partial_k K_{a b} \right),
\end{eqnarray} 
where $\simeq$ denotes equal to the principal part.  For brevity, we have
shown only the principal parts of the evolution equations as that is
what determines the hyperbolicity of the system.  The full evolution
equations are lengthy and available from the authors upon request.

We find that the eigenvalues of the characteristic matrix of the
system are $\{0,\pm 1,\pm c_1, \pm c_2, \pm c_3 \}$ where
\begin{eqnarray}
c_1 &=& \sqrt{2\sigma} \nonumber, \\
c_2 &=& \case{1}{2\sqrt{2}}\sqrt{\eta - 4 \eta \sigma - 2 \chi 
	- 12 \sigma \chi - 3 \eta \zeta} \nonumber, \\
c_3 &=& \case{1}{\sqrt{2}}\sqrt{2 + 4 \gamma - \eta - 2 \gamma \eta +
	2 \chi + 4 \gamma \chi - \eta \zeta}.
\end{eqnarray}

Thus in order for the system to be weakly hyperbolic, the parameters
must satisfy
\begin{eqnarray}
\sigma &\geq& 0 \nonumber, \\
\eta - 4 \eta \sigma - 2 \chi - 12 \sigma \chi - 3 \eta \zeta &\geq& 0 
\nonumber, \\
2 + 4 \gamma - \eta - 2 \gamma \eta + 2 \chi + 4 \gamma \chi - \eta \zeta
&\geq& 0.
\end{eqnarray}

If the above conditions are met, we find a complete set of eigenvectors,
so that the system is strongly hyperbolic, unless one of the following
conditions occur:

\begin{mathletters}
\begin{eqnarray}
c_i   &=& 0,\\
c_1 &=& c_3 \neq 1, \\
c_1 &=& c_3 = 1 \neq c_2. 
\end{eqnarray}
\end{mathletters}
If any of the above conditions are met, the system is only weakly
hyperbolic.  Note that if $\sigma=0$, then $c_1=0$, so that densitizing
the lapse is a necessary condition for strong hyperbolicty.  Also note
that if $\eta=\chi=0$, then $c_2=0$, so that constraints must be added
to the evolution equation for $d_{k i j}$ in order to have a strongly
hyperbolic system.

For physical characteristic speeds, each of the $c_i$ is either zero
or unity.  To make them all unity (the only choice that yields strongly
hyperbolic evolution equations) requires either
\begin{mathletters}
\begin{eqnarray}
\sigma &=&   1/2,\\
\zeta  &=&   -\frac{8 + 5\eta   + 10\gamma\eta}{\eta(7 + 6\gamma)},\\
\chi   &=&   -\frac{4 + 6\gamma - \eta - 3\gamma\eta}{(7 + 6\gamma)},
\end{eqnarray}
\end{mathletters}
or
\begin{equation}
\{ \sigma, \gamma, \zeta, \eta, \chi \} =  \{ \case{1}{2},
-\case{7}{6}, -\case{1}{9}(23+20\chi), \case{6}{5}, \chi \} .
\end{equation}

In the first case, there are two free parameters, and in the second
case there is one. In both cases, the evolution equations are strongly
hyperbolic as long as the free parameters are chosen such that all
five parameters are finite.

\subsection{Evolution of the constraints}
Taking ${\hat \partial_0}$ of the constraints, and replacing all
derivatives of the fundamental variables with the constraints and
their spatial derivatives, we obtain the following equations for the
evolution of the constraints:
\begin{eqnarray}
{\hat \partial_0} {\cal C} \simeq && 
- \case{1}{2} (2 - \eta + 2 \chi) N g^{p q} \partial_p {\cal C}_q, \\
{\hat \partial_0} {\cal C}_i \simeq && - (1 + 2 \gamma) N \partial_i {\cal C}
+ \case{1}{2} N g^{p q} g^{r s} \left[
(1 - \zeta) \partial_q {\cal C}_{p r s i} \right. \nonumber \\ && \mbox{}
\left. + (1 + \zeta) \partial_p {\cal C}_{s i q r}
- (1 + 2 \sigma) \partial_p {\cal C}_{q i r s} \right], \\
{\hat \partial_0} {\cal C}_{k i j} \simeq && 0, \\ 
{\hat \partial_0} {\cal C}_{k l i j} = && 
\case{1}{2} \eta N \left( g_{j [l} \partial_{k]} {\cal C}_i 
              + g_{i [l} \partial_{k]} {\cal C}_j \right)
 \nonumber \\ && \mbox{}
+ \chi N g_{i j} \partial_{[k} {\cal C}_{l]},
\end{eqnarray}
where again for brevity we have only shown the principal parts of
the equations.

The eigenvalues for the constraint evolution system are $\{0,\pm c_2,
\pm c_3 \}$. Because this is a subset of the eigenvalues of the
evolution equations, the constraints will propagate at the same speeds
as some of the characteristic fields of the evolved quantities.
Furthermore, we find that the constraint evolution system is strongly
hyperbolic whenever the regular evolution system is strongly
hyperbolic.

\subsection{Redefining the variables: System 2}
\label{sec:System2}
The evolution equations can also be modified by redefining the
variables that are evolved.  We define the generalized extrinsic
curvature $P_{i j}$ using the relation
\begin{equation}
\label{eq:Pdefinition}
P_{i j} \equiv K_{i j} + {\hat z} g_{i j} K,
\end{equation}
where ${\hat z}$ is an arbitrary parameter.  The inverse
transformation is given by
\begin{equation}
\label{eq:Pdefinitioninverse}
K_{i j} = P_{i j} + {\bar z} g_{i j} P,
\end{equation}
where $P \equiv g^{a b} P_{a b}$, and 
\begin{equation}
\label{eq:zbarhat}
{\bar z} = - \frac{\hat z}{1+3{\hat z}},
\end{equation}
which implies that ${\hat z} \neq - \case{1}{3}$ for the inverse 
transformation to exist.

We define the generalized derivative of the metric, $M_{k i j}$,
using the relation
\begin{eqnarray}
\label{eq:Mdefinition}
M_{k i j} = && \case{1}{2} \left\{ {\hat k} d_{k i j} + {\hat e} d_{(i
j) k} + g_{i j} \left[ {\hat a} d_k + {\hat b} b_k \right] 
\right. \nonumber \\ && \mbox{} \left. + g_{k (i}
\left[ {\hat c} d_{j)} + {\hat d} b_{j)} \right] \right\}.
\end{eqnarray}
The inverse transformation is given by
\begin{eqnarray}
\label{eq:Mdefinitioninverse}
d_{k i j} = && 2 \left\{ {\bar k} M_{k i j} + {\bar e} M_{(i j) k} +
g_{i j} \left[ {\bar a} M_k + {\bar b} W_k \right] 
\right. \nonumber \\ && \mbox{} \left. + g_{k (i} \left
[ {\bar c} M_{j)} + {\bar d} W_{j)} \right] \right\},
\end{eqnarray}
where the traces $M_k \equiv g^{a b} M_{k a b}$ and $W_k \equiv g^{a b}
M_{a b k}$, and 
\begin{mathletters}
\label{eq:barhattransformation}
\begin{eqnarray}
\delta {\bar a} = && 6 {\hat b}{\hat c}{\hat e} - 6 {\hat a}{\hat d}{\hat e} 
- {\hat a}{\hat e}^2 + {\hat b}{\hat e}^2 + {\hat c}{\hat e}^2
- {\hat d}{\hat e}^2 + 8 {\hat b}{\hat c}{\hat k} \nonumber \\ && \mbox{}
- 8 {\hat a}{\hat d}{\hat k} - 4 {\hat a}{\hat e}{\hat k} 
+ 2 {\hat b}{\hat e}{\hat k} + 2 {\hat c}{\hat e}{\hat k} 
- 4 {\hat a}{\hat k}^2,\\
\delta {\bar b} = && -8 {\hat b}{\hat c}{\hat e} + 8 {\hat a}{\hat d}{\hat e} 
+ 2 {\hat a}{\hat e}^2 - 2{\hat c}{\hat e}^2 - 4 {\hat b}{\hat c}{\hat k}
+ 4 {\hat a}{\hat d}{\hat k} \nonumber \\ && \mbox{}
+ 4 {\hat a}{\hat e}{\hat k} - 2 {\hat b}{\hat e}{\hat k} 
+ 2 {\hat d}{\hat e}{\hat k} - 4 {\hat b}{\hat k}^2,\\
\delta {\bar c} = && -8 {\hat b}{\hat c}{\hat e} +8 {\hat a}{\hat d}{\hat e} 
+ 2 {\hat a}{\hat e}^2 - 2 {\hat b}{\hat e}^2 - 4 {\hat b}{\hat c}{\hat k}
+ 4 {\hat a}{\hat d}{\hat k} \nonumber \\ && \mbox{}
+ 4 {\hat a}{\hat e}{\hat k} - 2 {\hat c}{\hat e}{\hat k} 
+ 2 {\hat d}{\hat e}{\hat k} - 4 {\hat c}{\hat k}^2,\\
\delta {\bar d} = && 4 {\hat b}{\hat c}{\hat e} - 4 {\hat a}{\hat d}{\hat e} 
- 4 {\hat a}{\hat e}^2 + 12 {\hat b}{\hat c}{\hat k}
- 12 {\hat a}{\hat d}{\hat k} + 4 {\hat b}{\hat e}{\hat k}  
\nonumber \\ && \mbox{}
+ 4 {\hat c}{\hat e}{\hat k} - 4 {\hat d}{\hat k}^2,\\
\delta_0 {\bar e} = && 2 {\hat e}, \\
\delta_0 {\bar k} = && - {\hat e} - 2 {\hat k}, \\
\delta_0 = && {\hat e}^2 - {\hat e}{\hat k} - 2 {\hat k}^2, \\
\delta = && \delta_0 (10 {\hat b}{\hat c} - 10 {\hat a}{\hat d} 
- {\hat a}{\hat e} + 3 {\hat b}{\hat e} + 3 {\hat c}{\hat e} 
+ {\hat d}{\hat e} + {\hat e}^2 \nonumber \\ && \mbox{}
- 6 {\hat a}{\hat k} - 2 {\hat b}{\hat k} - 2 {\hat c}{\hat k} 
- 4 {\hat d}{\hat k} - {\hat e}{\hat k} - 2 {\hat k}^2 ).
\end{eqnarray}
\end{mathletters}
For the inverse transformation to exist, $\delta \neq 0$.

Thus we have seven additional redefinition parameters $\{{\hat a},
{\hat b},{\hat c},{\hat d},{\hat e},{\hat k},{\hat z} \}$ (or
equivalently $\{{\bar a},{\bar b},{\bar c},{\bar d},{\bar e},{\bar
k},{\bar z} \}$) that can be used to modify the evolution equations.
Note that equations~(\ref{eq:zbarhat}) and
(\ref{eq:barhattransformation}) remain true under interchange of
$\{{\hat a}, {\hat b},{\hat c},{\hat d},{\hat e},{\hat k},{\hat z} \}$
and $\{{\bar a},{\bar b},{\bar c},{\bar d},{\bar e},{\bar k},{\bar z}
\}$.

When the principal terms in system 1 are transformed, terms containing
derivatives of the metric appear because of the traces
in~(\ref{eq:Pdefinitioninverse}) and~(\ref{eq:Mdefinitioninverse}).
These terms are eliminated using~(\ref{eq:dgdt})
and~(\ref{eq:definitiond}).

The redefinition parameters do not change the eigenvalues of the
evolution system, nor do they change whether or not the system is
strongly hyperbolic (see Appendix~\ref{sec:12paramhyper}).  In
addition, they have no effect on the principal part of the constraint
evolution equations.  The redefinition parameters, however, do affect
the eigenvectors of the evolution system and thus they also affect the
characteristic fields. In addition, the redefinition parameters change
the nonlinear terms in the nonprincipal parts of the evolution
equations and the constraint evolution system.

The principal parts of the evolution equations for $P_{i j}$ and $M_{k
i j}$ are 
\begin{eqnarray}
\widehat{\partial_0} g_{i j} \simeq && 0, \\
\widehat{\partial_0} P_{i j} \simeq && - N g^{a b} \left( 
 \mu_1 \partial_a M_{b i j}
+ \mu_2 \partial_a M_{(i j) b} 
+ \mu_3 \partial_{(i} M_{a b j)} \right. \nonumber \\ && \mbox{}
+ \mu_4 \partial_{(i} M_{j) a b} 
+ \mu_5 g_{i j} g^{c d} \partial_a M_{c d b} \nonumber \\ && \mbox{} \left.
+ \mu_6 g_{i j} g^{c d} \partial_a M_{b c d} \right), \\ 
\widehat{\partial_0} M_{k i j} \simeq && - N \left( 
\nu_1 \partial_k P_{i j}
+ \nu_2 \partial_{(i} P_{j) k}
+ \nu_3 g^{a b} g_{k (i} \partial_a P_{b j)}
\right. \nonumber \\ && \mbox{}
+ \nu_4 g_{i j} g^{a b} \partial_a P_{b k}
+ \nu_5 g^{a b} g_{k (i} \partial_{j)} P_{a b} \nonumber \\ && \mbox{} \left.
+ \nu_6 g_{i j} g^{a b} \partial_k P_{a b} \right), 
\end{eqnarray} 
where
\begin{mathletters}
\begin{eqnarray}
\mu_1 = && {\bar k} - \case{1}{2} (1 + \zeta) {\bar e}, \\
\mu_2 = && \case{1}{2} (1 - \zeta) {\bar e} - (1 + \zeta) {\bar k}, \\
\mu_3 = && (1 + 6 \sigma) {\bar b} - (1 - \zeta) {\bar k} 
- \case{1}{2} (1 - 4 \sigma - 3 \zeta) {\bar d} \nonumber \\ && \mbox{}
+ \case{1}{2} (1 + 4 \sigma + \zeta) {\bar e}, \\
\mu_4 = && (1 + 6 \sigma) {\bar a} + (1 + 2 \sigma) {\bar k} 
- \case{1}{2} (1 - 4 \sigma - 3 \zeta) {\bar c} \nonumber \\ && \mbox{}
- \case{1}{2} (1 - \zeta) {\bar e}, \\
\mu_5 = && (1 + 2 \gamma + 4 {\hat z} + 6 \gamma {\hat z} + 6 \sigma {\hat z})
{\bar b} - (\gamma + 2 {\hat z} + 3 \gamma {\hat z}) {\bar k} 
\nonumber \\ && \mbox{}
- \case{1}{2}(1 + 2 \gamma + 4 {\hat z} + 6 \gamma {\hat z} 
- 4 \sigma {\hat z} + \zeta) {\bar c}  \nonumber \\ && \mbox{}
+ \case{1}{2}(\gamma + 2 {\hat z} 
+ 3 \gamma {\hat z} + 4 \sigma  {\hat z}) {\bar e}, \\
\mu_6 = && (1 + 2 \gamma + 4 {\hat z} + 6 \gamma {\hat z} + 6 \sigma {\hat z})
{\bar a}   \nonumber \\ && \mbox{}
+ (\gamma + 2 {\hat z} + 3 \gamma {\hat z} + 2 \sigma {\hat z})
{\bar k}   \nonumber \\ && \mbox{}
- \case{1}{2}(1 + 2 \gamma + 4 {\hat z} + 6 \gamma {\hat z} 
- 4 \sigma {\hat z} + \zeta) {\bar d}  \nonumber \\ && \mbox{}
- \case{1}{2}(\gamma + 2 {\hat z} 
+ 3 \gamma {\hat z}) {\bar e}, \\
\nu_1 = && {\hat k}, \\
\nu_2 = && {\hat e}, \\
\nu_3 = && \case{1}{2}(2 - 2 \eta - \chi) {\hat d} 
- \case{1}{2}(\eta + 3 \chi) {\hat c}
- \case{1}{4}(\eta + 2 \chi) {\hat e}  \nonumber \\ && \mbox{}
- \case{1}{2} \eta {\hat k}, \\
\nu_4 = && \case{1}{2}(2 - 2 \eta - \chi) {\hat b} 
- \case{1}{2}(\eta + 3 \chi) {\hat a}
- \case{1}{4} \eta {\hat e}
- \case{1}{2} \chi {\hat k}, \\
\nu_5 = && \case{1}{2}(2 + \eta + 3 \chi + 6 {\bar z} + 2 \eta {\bar z}
+ 6 \chi {\bar z}) {\hat c}   \nonumber \\ && \mbox{}
+ \case{1}{2}(2 \eta + \chi + 2 {\bar z} + 4 \eta {\bar z}
+ 2 \chi {\bar z}) {\hat d}  
+ \case{1}{2}(\eta + 2 \eta {\bar z}) {\hat k} \nonumber \\ && \mbox{}
+ \case{1}{4}(\eta + 2 \chi + 4 {\bar z} + 2 \eta {\bar z}
+ 4 \chi {\bar z}) {\hat e}, \\
\nu_6 = && \case{1}{2}(2 + \eta + 3 \chi + 6 {\bar z} + 2 \eta {\bar z}
+ 6 \chi {\bar z}) {\hat a}   \nonumber \\ && \mbox{}
+ \case{1}{2}(2 \eta + \chi + 2 {\bar z} + 4 \eta {\bar z}
+ 2 \chi {\bar z}) {\hat b} \nonumber \\ && \mbox{}
+ \case{1}{4}(\eta + 2 \eta {\bar z}) {\hat e}
+ \case{1}{2}(\chi + 2 {\bar z} + 2 \chi {\bar z}) {\hat k}.
\end{eqnarray}
\end{mathletters}
Again, the full evolution equations are available from the authors
upon request.  

Furthermore, we note that if $\mu_i = \kappa \nu_i$ for all $i$ and
constant $\kappa$, the system is symmetrizable hyperbolic using the energy
norm argument of \cite{frittelli_reula96}.  These conditions, however,
do not have to be met for the system to be well-posed.  It is possible
to construct a symmetrizer for any of the strongly hyperbolic systems.

\subsection{Evolving with contravariant indices}
So far, we have written all of our fundamental variables with
covariant indices.  Alternatively, we could have defined the new variable
\begin{equation}
D_k^{~ i j} \equiv \partial_k g^{i j}.
\end{equation}
Note that $d_{k i j} = - D_{k i j}$.  If we evolve $\{g^{i j},
P^{i j}, M_k^{~ i j} \}$ instead of $\{g_{i j}, P_{i j}, M_{k i j} \}$,
it would result in only trivial changes to the principal parts of the
equations.  The characteristic speeds would be unchanged, as would the
nature of the hyperbolicity of the system, since the principal part of
the metric evolution equation is zero (See
Appendix~\ref{sec:12paramhyper}).  The only changes would occur in the
nonlinear terms of the evolution equations.

\subsection{Frittelli-Reula system}
We recover the system of \cite{frittelli_reula96} if we make the following
choices for our parameters:
\begin{mathletters}
\label{eq:FRparams}
\begin{eqnarray}
\{ \sigma, \gamma, \zeta, \eta, \chi \} &= & 
\{ - {\bar \epsilon} \case{1 + 3 {\bar \alpha}}{2} ,
\case{2 {\bar \gamma}}{1 + 3 {\bar \beta}},1,4,
\case{-4 {\bar \alpha}}{1 + 3 {\bar \alpha}}\}, \\
\{ {\hat z}, {\hat k}, {\hat a}, {\hat b}, {\hat c}, {\hat d},
{\hat e} \}   &=& \{{\bar \beta},1,{\bar \alpha},0,0,0,0\},
\end{eqnarray}
\end{mathletters}
where $\{{\bar \alpha},{\bar \beta},{\bar \gamma},{\bar \epsilon}\}$
correspond to $\{\alpha,\beta,\gamma,\epsilon\}$ in
\cite{frittelli_reula96}. However, as pointed out in
\cite{Stewart1998}, this system is not symmetric hyperbolic unless 
the term $-2h^{l (i} M^{j) k}_{~ ~ ~ l,k}$ in Eq. (16) of
\cite{frittelli_reula96} is replaced with $-2h^{l (i} M^{j) k}_{~ ~ ~
k,l}$ by adding a term proportional to the
constraint~(\ref{eq:fic}).
In our system this corresponds to
changing $\zeta=1$ to $\zeta=-1$ in~(\ref{eq:FRparams}).  

In \cite{frittelli_reula94,Stewart1998,Brodbeck1999}, this correction
has been made for the parameter choice $\{{\bar \alpha},{\bar
\beta},{\bar \gamma},{\bar \epsilon}\} = \{-1,-1,1,\case{1}{2}\}$; we
recover this system if we choose our parameters to be
\begin{mathletters}
\begin{eqnarray}
\{ \sigma, \gamma, \zeta, \eta, \chi \} &= & \{ \case{1}{2},-1,-1,4,-2 \}, \\
\{ {\hat z}, {\hat k}, {\hat a}, {\hat b}, {\hat c}, {\hat d},
{\hat e} \}   &=& \{-1,1,-1,0,0,0,0\}.
\end{eqnarray}
\end{mathletters}

The system of \cite{frittelli_reula94,Stewart1998,Brodbeck1999}
was further generalized by \cite{Hern1999}, who used the
constraints to modify the evolution equations in a manner similar
to that in
Sec.~\ref{sec:System1}.  We recover the system of \cite{Hern1999} by choosing:
\begin{mathletters}
\begin{eqnarray}
\{ \sigma, \gamma, \zeta, \eta, \chi \} &= & 
\{ \case{1}{2},-{\tilde \gamma},2{\tilde \Theta}-1,4 {\tilde \eta},
-2 {\tilde \eta} \}, \\
\{ {\hat z}, {\hat k}, {\hat a}, {\hat b}, {\hat c}, {\hat d},
{\hat e} \}   &=& \{-1,1,-1,0,0,0,0\},
\end{eqnarray}
\end{mathletters}
where $\{{\tilde \gamma},{\tilde \Theta},{\tilde \eta}\}$ correspond
to $\{\gamma,\Theta,\eta\}$ in \cite{Hern1999}.

\subsection{Einstein-Christoffel system}
\label{sec:Einst-Christoffel-sy}
We recover the system of \cite{anderson_york99}  if we make the following
choices for our parameters:
\begin{mathletters}
\begin{eqnarray}
\{ \sigma, \gamma, \zeta, \eta, \chi \} &= & \{ \case{1}{2},0,-1,4,0 \}, \\
\{ {\hat z}, {\hat k}, {\hat a}, {\hat b}, {\hat c}, {\hat d},
{\hat e} \}   &=& \{0,1,0,0,2,-2,0\}.
\end{eqnarray}
\end{mathletters}
This system is symmetrizable hyperbolic and has a very simple principal part
\begin{mathletters}
\label{eq:ecprincipalpart}
\begin{eqnarray}
\widehat{\partial_0} P_{i j} \simeq && 
- N g^{a b} \partial_a M_{b i j} ,\\
\widehat{\partial_0} M_{k i j} \simeq && 
- N \partial_k P_{i j}.
\end{eqnarray}
\end{mathletters}
Essentially this system is a set of
six (one for each $\{i,j\}$ pair) coupled quasilinear scalar wave
equations with nonlinear source terms.

\subsection{Generalized Einstein-Christoffel: System 3}
If we examine the principal part of System 2, and demand that $\mu_1 =
\nu_1 = 1$ and all other $\mu_i$ and $\nu_i$ vanish, we obtain a
two-parameter system $\{\eta,{\hat z}\}$ that has the same simple
wave-like form~(\ref{eq:ecprincipalpart}) as the Einstein-Christoffel
system.  This system is obtained by setting
\begin{mathletters}
\begin{eqnarray}
\{ \sigma, \gamma, \zeta, \eta, \chi \} &= & 
\{ \case{1}{2},\case{-4 + \eta}{2 \eta},-1,\eta, \case{-4 + \eta}{4} \}, \\
\{ {\hat z}, {\hat k}, {\hat a}, {\hat b}, {\hat c}, {\hat d},
{\hat e} \}   &=& \{{\hat z},1,
\case{-4 + \eta - 12 {\hat z} + 9 \eta {\hat z}}{2 \eta},\nonumber \\ &&
\case{4 - \eta + 12 {\hat z} - 7 \eta {\hat z}}{2 \eta}, 
2,-2,0\},
\end{eqnarray}
\end{mathletters}
where ${\hat z} \neq -\case{1}{3}$ and $\eta \neq 0$.  This system has
physical characteristic speeds and is symmetrizable hyperbolic.  The
free parameter $\eta$ will affect the principal part of the constraint
evolution equations, while the parameter ${\hat z}$ will affect only
the nonlinear terms in the evolution equations and the constraint
evolution equations.  It is this system that we will explore
numerically in the following section.  The complete equations for this
system are available upon request from the authors.

The characteristic eigenfields of this system are particularly simple,
and can be obtained from~(\ref{eq:ecprincipalpart}) without the use of
the lengthy decomposition procedure described in
Appendix~\ref{sec:charsystem}. In a direction $\xi_i$, the eigenfields
are
\begin{mathletters}
\begin{eqnarray}
U^{0}_{ij}  &\equiv& g_{ij},\\
U^{0}_{kij} &\equiv& M_{kij} - \xi_k\xi^\ell M_{\ell ij},\\
U^{\pm}_{ij}&\equiv& P_{ij} \pm \xi^k M_{kij}.
\end{eqnarray}
\end{mathletters}
The $U^0$ quantities propagate along the normal to the time slice 
(coordinate speed $-\beta^i$), and the $U^{\pm}$ quantities propagate
along the light cone (coordinate speed $-\beta^i\pm N\xi^i$).

\section{Numerical Results}
\label{sec:Numerical-Results}
In this section we present results from a numerical code that solves
the evolution equations of System 3 in three spatial dimensions plus
time. This code, which will be described in detail
elsewhere\cite{Kidder2001a}, is a three-dimensional generalization of
a spherically symmetric code discussed previously\cite{Kidder2000a},
and is based on pseudospectral collocation methods.  Our code works
in full 3D; we do not exploit any symmetries of the black hole
solutions that we evolve.

In this paper, we will concern ourselves only with single black hole
spacetimes.  In this case, we solve the evolution equations in a
spherical shell extending from inside the horizon to some artificial
outer boundary. Although we use standard spherical polar coordinates
$({r,\theta,\phi})$, we evolve the {\it Cartesian\/} components of our
variables; this allows us to use scalar spherical harmonics $Y_{\ell
m}(\theta,\phi)$ as angular basis functions for all quantities. We
use Chebyshev polynomials as the basis functions in radius.  

As described in\cite{Kidder2000a}, we use the method of lines in order
to integrate forward in time with a fourth-order Runge-Kutta method.
Boundary conditions are imposed by constructing the characteristic
fields that propagate normal to the boundary, and imposing conditions
only on those fields that propagate into the computational domain.
Since all characteristic fields at the inner boundary are outgoing
(into the hole), no boundary condition is needed there and none is
imposed. At the outer boundary, we impose $\partial_t U^{-} = 0$ on
each of the characteristic fields $U^-$ that is ingoing there.  We use
analytic initial data corresponding to time-independent slicings of a
single black hole, and fix the gauge quantities $Q$ and $\beta^i$ to
their analytic values.  Note that the constraint equations are not
solved explicitly, but are instead used as a check on the accuracy of
our numerical integrations.

\subsection{Einstein-Christoffel}
Figure~\ref{fig:HW03-05-2001_ECMxl2} shows the $\ell_2$ norm of a
component of the momentum constraint
for several evolutions of a Schwarzschild black hole using the
Einstein-Christoffel system, which is equivalent to System 3 with
$\eta=4$ and $\hat{z}=0$.  Initially the fields are given analytically
on a Painlev\'{e}-Gullstrand time
slice\cite{Gundlach1999,painleve21,gullstrand22,Martel2000}. Explicit
formulae for our variables on the initial slice can be found
in~\cite{Kidder2000a}.

As is evident from Figure~\ref{fig:HW03-05-2001_ECMxl2}, the
constraint increases with time until the simulation terminates.  The
evolutions with higher radial resolution run longer, but increase at
approximately the same rate.  In addition, for a fixed resolution, we
see no significant dependence on $\Delta t$, and for a fixed radial
resolution and time step, we see no significant dependence on angular
resolution. This
suggests that the growth of the constraints may be due to an
unphysical solution of the equations rather than a numerical
instability.  Numerical instabilities typically become worse when one
increases the resolution or decreases the time step.  In contrast,
our results appear consistent with an unphysical solution of the equations
that initially has a nonzero amplitude because of small numerical errors.

\subsection{Generalized Einstein-Christoffel}
\begin{figure}
\begin{center}
\begin{picture}(240,240)
\put(0,0){\epsfxsize=3.5in\epsffile{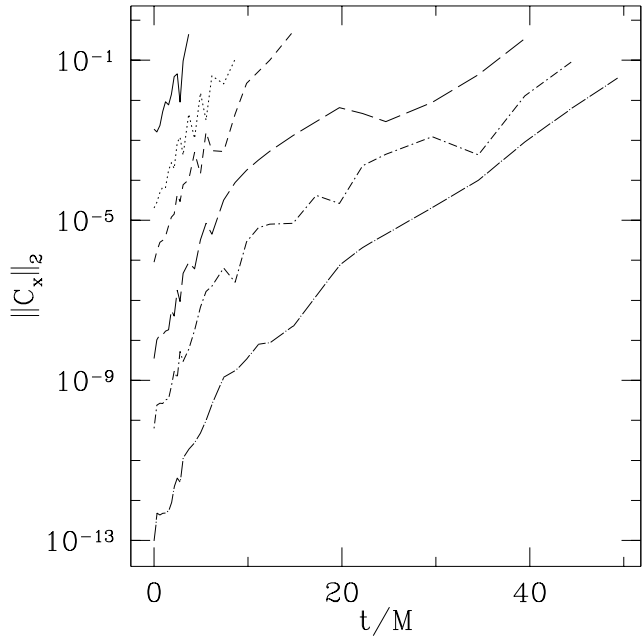}}
\end{picture}
\end{center}
\caption{Momentum constraint ${\cal C}_x$ versus time for evolutions
of a Painlev\'{e}-Gullstrand time slicing of a Schwarzschild black
hole using the Einstein-Christoffel system. Results are plotted for
several radial resolutions ranging from $N_r=10$ to $N_r=40$, fixed
angular resolution $\ell = 7$ and fixed time resolution $\Delta t =
0.015M$. Higher radial resolutions correspond to smaller errors.}
\label{fig:HW03-05-2001_ECMxl2}
\end{figure}

\begin{figure}
\begin{center}
\begin{picture}(240,240)
\put(0,0){\epsfxsize=3.5in\epsffile{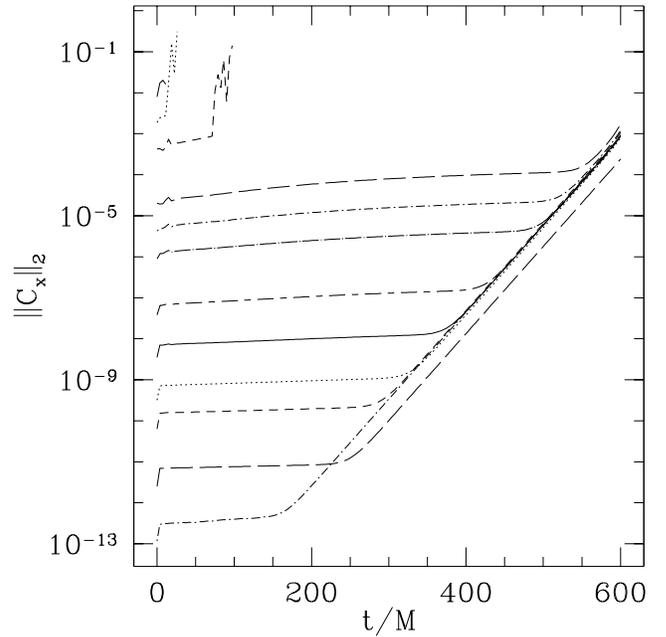}}
\end{picture}
\end{center}
\caption{Momentum constraint ${\cal C}_x$ versus time for 
the same evolutions shown in Figure~\ref{fig:HW03-05-2001_ECMxl2}
except $\eta=4/33$ and $\hat{z}=-1/4$, and we plot more radial resolutions.
If the outer boundary is moved out to $r=40M$, the run time extends to
$\sim 1300M$ for the same accuracy.}
\label{fig:LKWF09-19-2000A_ECMxl2}
\end{figure}

Because we suspected that the instability shown in
Figure~\ref{fig:HW03-05-2001_ECMxl2} is related to the equations
rather than the numerical method, we repeated the above evolutions for
various values of the free parameters $\eta$ and $\hat{z}$, searching
the two-dimensional parameter space for systems of evolution equations
that might be better-behaved.  We found that for $\eta\simeq 4/33$ and
$\hat{z}\simeq -1/4$, our numerical simulations ran for an order of
magnitude longer than for the basic Einstein-Christoffel
system. Typical results are plotted in
Figure~\ref{fig:LKWF09-19-2000A_ECMxl2}.  Although a growing mode is
still present, its growth rate is much smaller than in
Figure~\ref{fig:HW03-05-2001_ECMxl2}, and the momentum constraint
is less than $10^{-3}$ until approximately $600M$.

We see no evidence that the growth is due to a numerical
instability. In contrast, the evolutions in
Figure~\ref{fig:LKWF09-19-2000A_ECMxl2} appear to converge to a
well-defined solution. 
This solution is the sum of two components: a roughly time-independent
component and an exponentially growing component. By extrapolating
backwards along the growing component in
Figure~\ref{fig:LKWF09-19-2000A_ECMxl2}, one can see that this
component has magnitude $\sim10^{-16}$ at $t=0$, which is on the order
of machine roundoff error. 

As in the Einstein-Christoffel case, we see no dependence on angular
resolution or on $\Delta t$.  Our results do depend upon the location
of the outer boundary.  In the evolution shown in
Figure~\ref{fig:LKWF09-19-2000A_ECMxl2}, the spherical domain extends
from $r=1.9M$ to $r=11.9M$.  Moving the outer boundary further out
results in longer evolutions, increasing the run time from around
$600M$ up to $1300M$ with the outer boundary at $r=40M$.
Moving the outer boundary beyond $r=40M$, however, does not have any
effect. We suspect that errors in the constraints at the outer
boundary may be responsible for the constraint-violating modes.

In addition to Painlev\'{e}-Gullstrand slicings, we have run
Kerr-Schild\cite{marsa96,Matzner_Huq1999}
and harmonic-time\cite{bona_masso88,cook_scheel97} slicings of a Schwarzschild
black hole with similar qualitative results.  For example, using the
parameters of Figure~\ref{fig:LKWF09-19-2000A_ECMxl2} with a
Kerr-Schild slicing as initial data, we were able to evolve up to
$t=500M$ with the outer boundary at $r=11.9M$, and up to $t=900M$ with
the outer boundary at $r=40M$.  We have also evolved a Kerr black hole
with $a=M/2$ to $t=400M$, with a spherical shell extending
from $r=1.5M$ to $r=11.5M$.

\section{Discussion}
\label{sec:Discussion}
We have constructed a twelve-parameter family of hyperbolic
formulations of Einstein's equations that is strongly hyperbolic
for a wide range of the parameter space, and that includes the systems
of \cite{frittelli_reula96} and \cite{anderson_york99}.  By restricting
ourselves to a two-parameter subset of these equations, we have
demonstrated how the choice of parameters can have a dramatic effect
upon the amount of time a numerical simulation of a black hole can run
before being swamped by an unphysical solution.  

Our runs with our best parameter choices appear to be limited only by
the growth of constraint-violating modes which grow from the level of
numerical roundoff errors.  At present, we have no explanation as to
why the particular choice of parameters used to produce
Figure~\ref{fig:LKWF09-19-2000A_ECMxl2} is so much better than the
Einstein-Christoffel system.  This choice was found empirically by
running our code for various values of the parameters.  It would be
extremely useful to have some theoretical understanding of why one
particular parameter choice behaves much better than another, as the
cost of performing a parameter search on the full twelve-parameter
system would be prohibitive.

Having found a system of equations and a numerical method capable of
evolving a single black hole for a physically interesting length of
time, we now plan to turn our attention to the evolution of a
binary black hole system.  We expect our computational method to be
capable of evolving the binary system to times on the order of several
hundred $M$ given appropriate gauge conditions.  When we realize this, 
we will be able to simulate the last orbit or two prior to the plunge
as well as the coalescence itself.

\acknowledgements 

We thank Harald Pfeiffer, Manuel Tiglio, and James W. York, Jr. for
helpful discussions.  This work was supported in part by NSF grants
PHY-9900672 and PHY-0084729 and NASA grant NAG5-7264.  Computations
were performed on the National Computational Science Alliance SGI
Origin 2000, and on the Wake Forest University Department of Physics
IBM SP2 with support from an IBM SUR grant.

\appendix

\section{Hyperbolicity}
\label{sec:charsystem}
To determine the characteristic speeds and eigenvectors of a system of
the form~(\ref{eq:firstorderform}), we proceed in two steps. Instead
of directly finding the eigenvalues and eigenvectors of $C \equiv A^i
\xi_i$, we first construct a transformation $D$ such that $C' \equiv D
A^i \xi_i D^{-1}$ is independent of the direction $\xi_i$ and of the
metric quantities $g_{ij}$. We then solve $C' w_i = \lambda_i w_i$.
The eigenvalues of the original matrix $C$ are $\lambda_i$, and the
eigenvectors are $D^{-1} w_i$.

The transformation $D$ is the decomposition of each of the fundamental
tensor (or tensor-like) quantities into its irreducible parts, as we
now describe.  Suppose $v \equiv D u$.  Then if $u$ and $v$ are
scalars, $D$ is the identity operator, $v = u$.  For a vector quantity
$u = V_i$, $D$ is defined by
\begin{equation}
V_i =  D^{-1} v = V^{(T)}_i + \xi_i V^{(L)},
\end{equation}
where the longitudinal and transverse parts of $V_i$ are given by
\begin{mathletters}
\begin{eqnarray}
V^{(L)}    &\equiv& \xi^m V_m,\\
V^{(T)}_i  &\equiv& \perp^{~ m}_i V_m,
\end{eqnarray}
\end{mathletters}
where $\perp_{ij}$ is the projection operator
\begin{equation}
\perp_{i j}\equiv g_{i j} - \xi_i \xi_j.
\end{equation}

For a symmetric second-rank tensor $u = P_{ij}$,
\begin{eqnarray}
P_{i j} &=& D^{-1} v \nonumber \\ 
&=& P^{(TTs)}_{i j} + 2 \xi_{(i} P^{(LT)}_{j)}
+ \case{1}{2}(3 \xi_i \xi_j - g_{i j}) P^{(LL)}
\nonumber \\ && \mbox{} + \case{1}{2}(g_{i j} -\xi_i \xi_j) P,
\end{eqnarray}
where
\begin{mathletters}
\begin{eqnarray}
P              &\equiv& g^{m n} P_{m n},\\
P^{(LL)}       &\equiv& \xi^m \xi^n P_{m n},\\
P^{(LT)}_i     &\equiv& \xi^m \perp_i^{~ n} P_{m n},\\
P^{(TTs)}_{i j} &\equiv& \left(\perp^{~ ~ m}_{(i} \perp^{~ ~ n}_{j)}
- \case{1}{2} \perp_{i j} \perp^{m n} \right) P_{m n}.
\end{eqnarray}
\end{mathletters}

For a third-rank object $u = M_{k i j}$, symmetric on its last two
indices,
\begin{eqnarray}
\label{eq:mdecomp}
M_{k i j} &=& D^{-1} v \nonumber \\ 
&=& M^{(TTT)}_{k i j} + 2 \xi_{(i} M^{(TTLs)}_{j) k}
+ 2 \xi_{(i} M^{(TTLa)}_{j) k} \nonumber \\ && \mbox{}
+ \xi_k M^{(LTT)}_{i j}
+ \case{1}{4} M^{(TLL)}_k \left( 7 \xi_i \xi_j - 3 g_{i j} \right)
\nonumber \\ && \mbox{}
+ \case{1}{2} M^{(TLL)}_{(i} \left( g_{j) k} - \xi_{j)} \xi_k \right)
\nonumber \\ && \mbox{}
+ \case{1}{2} M^{(LLT)}_k \left( g_{i j} - \xi_i \xi_j \right)
\nonumber \\ && \mbox{}
+ M^{(LLT)}_{(i} \left( 3 \xi_{j)} \xi_k - g_{j) k} \right)
\nonumber \\ && \mbox{}
+ \case{3}{4} M^{(TRR)}_k \left( g_{i j} - \xi_i \xi_j \right)
\nonumber \\ && \mbox{}
+ \case{1}{2} M^{(TRR)}_{(i} \left( \xi_{j)} \xi_k - g_{j) k} \right)
\nonumber \\ && \mbox{}
+ \case{1}{2} M^{(RRT)}_k \left( \xi_i \xi_j - g_{i j} \right)
\nonumber \\ && \mbox{}
+ M^{(RRT)}_{(i} \left( g_{j) k} - \xi_{j)} \xi_k \right)
\nonumber \\ && \mbox{}
+ \case{1}{2} M^{(LLL)} \left( 5 \xi_k \xi_i \xi_j - \xi_k g_{i j}
- 2 g_{k (i} \xi_{j)} \right)
\nonumber \\ && \mbox{}
+ \case{1}{2} M^{(LRR)} \left( \xi_k g_{i j} -  \xi_k \xi_i \xi_j \right)
\nonumber \\ && \mbox{}
+ M^{(RRL)} \left( g_{k (i} \xi_{j)} - \xi_k \xi_i \xi_j\right),
\end{eqnarray}
where
\begin{mathletters}
\begin{eqnarray}
M^{(RRL)} &\equiv& g^{c a} \xi^b M_{c a b},\\
M^{(LRR)} &\equiv& g^{a b} \xi^c M_{c a b},\\
M^{(LLL)} &\equiv& \xi^c \xi^a \xi^b M_{c a b},\\
M^{(RRT)}_i &\equiv& g^{c a} \perp_i^{~ b} M_{c a b},\\
M^{(TRR)}_i &\equiv& g^{a b} \perp_i^{~ c} M_{c a b},\\
M^{(LLT)}_i &\equiv& \xi^c \xi^a \perp_i^{~ b} M_{c a b},\\
M^{(TLL)}_i &\equiv& \xi^a \xi^b \perp_i^{~ c} M_{c a b},\\
M^{(LTT)}_{i j} &\equiv& \xi^c \left(\perp^{~ a}_i \perp^{~ b}_j
- \case{1}{2} \perp_{i j} \perp^{a b} \right) M_{c a b},\\
M^{(TTLs)}_{i j} &\equiv& \xi^b \left(\perp^{~ ~ c}_{(i} \perp^{~ ~ a}_{j)}
- \case{1}{2} \perp_{i j} \perp^{c a} \right) M_{c a b},\\
M^{(TTLa)}_{i j} &\equiv& \xi^b \perp^{~ ~ c}_{[i} \perp^{~ ~ a}_{j]} 
M_{c a b},\\
M^{(TTT)}_{k i j} &\equiv& \left[ \perp_k^{~ c} \perp_i^{~ a} \perp_j^{~ b}
\right. \nonumber \\ && \mbox{} 
- \case{1}{4} \perp_{i j} \left( 3 \perp^{a b} \perp_k^{~ c} 
- \perp^{c a} \perp_k^{~ b} - \perp^{c b} \perp_k^{~ a} \right)
\nonumber \\ && \mbox{} 
- \case{1}{4} \perp_{k j} \left( 3 \perp^{c b} \perp_i^{~ a} 
- \perp^{c a} \perp_i^{~ b} - \perp^{a b} \perp_i^{~ c} \right)
\nonumber \\ && \mbox{} 
- \case{1}{4} \perp_{k i} \left( 3 \perp^{c a} \perp_j^{~ b} 
- \perp^{a b} \perp_j^{~ c} \right. \nonumber \\ && \mbox{} \left. \left.
- \perp^{c b} \perp_j^{~ a} \right)
\right] M_{c a b}.
\end{eqnarray}
\end{mathletters}

Finally for a four-index object ${\cal C}_{k l i j}$, symmetric on its last 
two indices and antisymmetric on its first two indices,
\begin{eqnarray}
\label{eq:ficdecomp}
{\cal C}_{k l i j} &=& D^{-1} v \nonumber \\ 
&=& {\cal C}^{(TF)}_{k l i j} 
+ \case{3}{5} {\cal C}^{(TTRRa)}_{k l} g_{i j}
+ \case{4}{5} {\cal C}^{(RTTRa)}_{k l} g_{i j}
\nonumber \\ && \mbox{}
+ \case{2}{5} \left( {\cal C}^{(TTRRa)}_{i [k} g_{l] j} 
+ {\cal C}^{(TTRRa)}_{j [k} g_{l] i} \right)
\nonumber \\ && \mbox{}
+ \case{6}{5} \left( {\cal C}^{(RTTRa)}_{i [k} g_{l] j} 
+ {\cal C}^{(RTTRa)}_{j [k} g_{l] i} \right)
\nonumber \\ && \mbox{}
- \case{2}{3} \left( {\cal C}^{(RTTRs)}_{i [k} g_{l] j} 
+ {\cal C}^{(RTTRs)}_{j [k} g_{l] i} \right)
\nonumber \\ && \mbox{}
- \case{6}{5} {\cal C}^{(LTRR)}_{[k} \xi_{l]} g_{i j}
+ \case{4}{5} {\cal C}^{(LTRR)}_{[k} g_{l] (j} \xi_{i)} 
\nonumber \\ && \mbox{}
+ \case{4}{5} {\cal C}^{(LTRR)}_{(i} g_{j) [k} \xi_{l]} 
- \case{4}{5} {\cal C}^{(RLTR)}_{[k} \xi_{l]} g_{i j}
\nonumber \\ && \mbox{}
+ \case{8}{15} {\cal C}^{(RLTR)}_{[k} g_{l] (j} \xi_{i)} 
+ \case{28}{15} {\cal C}^{(RLTR)}_{(i} g_{j) [k} \xi_{l]}
\nonumber \\ && \mbox{}
+ \case{4}{5} {\cal C}^{(RTLR)}_{[k} \xi_{l]} g_{i j}
- \case{28}{15} {\cal C}^{(RTLR)}_{[k} g_{l] (j} \xi_{i)} 
\nonumber \\ && \mbox{}
- \case{8}{15} {\cal C}^{(RTLR)}_{(i} g_{j) [k} \xi_{l]} 
+ 2 \xi_{(i} g_{j) [k} \xi_{l]} {\cal C}^{(RLLR)},
\end{eqnarray}
where
\begin{mathletters}
\begin{eqnarray}
{\cal C}^{(RLLR)} &\equiv& g^{c b} \xi^d \xi^a {\cal C}_{c d a b},\\
{\cal C}^{(RTLR)}_i &\equiv& g^{c b} \xi^a \perp_i^{~ d} {\cal C}_{c d a b},\\
{\cal C}^{(RLTR)}_i &\equiv& g^{c b} \xi^d \perp_i^{~ a} {\cal C}_{c d a b},\\
{\cal C}^{(LTRR)}_i &\equiv& g^{a b} \xi^c \perp_i^{~ d} {\cal C}_{c d a b},\\
{\cal C}^{(RTTRs)}_{i j} &\equiv& g^{c b} 
\left(\perp^{~ ~ d}_{(i} \perp^{~ ~ a}_{j)}
- \case{1}{2} \perp_{i j} \perp^{d a} \right) {\cal C}_{c d a b},\\
{\cal C}^{(RTTRa)}_{i j} &\equiv& g^{c b} \perp^{~ ~ d}_{[i} 
\perp^{~ ~ a}_{j]} {\cal C}_{c d a b},\\
{\cal C}^{(TTRRa)}_{i j} &\equiv& g^{a b} \perp^{~ ~ c}_{[i} 
\perp^{~ ~ d}_{j]} {\cal C}_{c d a b},\\
{\cal C}^{(TF)}_{k l i j} &\equiv& \left(
g_k^{~ c} g_l^{~ d} g_i^{~ a} g_j^{~ b} 
- \case{28}{15} g^{c b} g^a_{~ (i} g_{j) [k} g_{l]}^{~ ~ d}
\right. \nonumber \\ && \mbox{}
+ \case{8}{15} g^{c b} g^d_{~ (i} g_{j) [k} g_{l]}^{~ ~ a}
- \case{4}{5} g^{a b} g^d_{~ (i} g_{j) [k} g_{l]}^{~ ~ c}
\nonumber \\ && \mbox{} 
- \case{4}{5} g_{i j} g^{c b} g^d_{~ [k} g_{l]}^{~ ~ a}
\nonumber \\ && \mbox{} \left.
- \case{3}{5} g_{i j} g^{a b} g_k^{~ c} g_l^{~ d}
\right) {\cal C}_{c d a b}.
\end{eqnarray}
\end{mathletters}
Strictly speaking, eqs.~(\ref{eq:mdecomp}) and~(\ref{eq:ficdecomp})
are not complete irreducible decompositions.  However, they are
sufficient for our purposes.

If $u$ consists of several tensor (or tensor-like) objects, then the
effect of $D$ is to transform each object independently according to
the above definitions. In matrix language, this means that $D$ is
block diagonal.

\section{Change of variables and hyperbolicity}
\label{sec:12paramhyper}
In this section we show that for a system of the
form~(\ref{eq:firstorderform}), a change of variables (such as the
transformation from system 1 to system 2, or the raising and lowering of
tensor indices of fundamental variables) does not change either the
characteristic speeds or whether the system is strongly hyperbolic, provided
that the following conditions are met:
\begin{enumerate}
	\item The change of variables is linear in all dynamical variables
	      except possibly the metric.
	\item The change of variables is invertible.
	\item Time and space derivatives of the metric can be written
	      as a sum of only non-principal terms (for example,
	      using~(\ref{eq:dgdt}) and~(\ref{eq:definitiond})).
\end{enumerate}

For a system of the form~(\ref{eq:firstorderform}), we choose an
arbitrary direction $\xi_i$ and we define the matrix $C$ according
to~(\ref{eq:characteristicmatrix}).  The system has $k$ characteristic
speeds $\lambda^{(k)}$ and eigenvectors $w^{(k)}$ that obey
\begin{equation}
\label{eq:eigenvalueequation}
C w^{(k)} = \lambda^{(k)} w^{(k)}.
\end{equation}
If $M$ is the matrix whose columns are the eigenvectors $w^{(k)}$, then
strong hyperbolicity is equivalent to $\det M \neq 0$, with all
$\lambda^{(k)}$ real.

Now consider a change of variables $v = T u$, where $T$ is
a matrix. If we multiply~(\ref{eq:firstorderform}) on the left by $T$, we
obtain
\begin{eqnarray}
\label{eq:firstorderformtransformed}
\hat{\partial}_0 v + T A^i T^{-1} \partial_i v 
&=& T F+(\hat{\partial}_0 T) u + T A^i T^{-1} (\partial_i T) u \nonumber \\
&=& F'.
\end{eqnarray}
In the last step, we have used property 1 above to rewrite $\partial_i T$
and $\hat{\partial}_0 T$ in terms of derivatives of the metric, and
we have used property 3 to eliminate these derivatives, absorbing
the resulting non-principal terms into the new right-hand side $F'$.

The characteristic matrix for~(\ref{eq:firstorderformtransformed})
in the direction $\xi_i$ is $C' \equiv T A^i T^{-1} \xi_i$. 
Note that
\begin{equation}
C' T w^{(k)} =  T A^i T^{-1} \xi_i T w^{(k)} = 
T A^i \xi_i w^{(k)} = \lambda^{(k)} T w^{(k)},
\end{equation}
so~(\ref{eq:firstorderformtransformed}) and~(\ref{eq:firstorderform})
have the same characteristic speeds $\lambda^{(k)}$, and the eigenvectors
of~(\ref{eq:firstorderformtransformed}) are $T w^{(k)}$.

Furthermore, the matrix of eigenvectors 
for~(\ref{eq:firstorderformtransformed}) is $M'=(TM)^{T}$, so
\begin{equation}
\det M' = \det (TM)^{T} = \det T \det M.
\end{equation}
If the transformation $T$ is invertible, $\det M' \neq 0$
if and only if $\det M \neq 0$ , so~(\ref{eq:firstorderformtransformed})
is strongly hyperbolic if and only if~(\ref{eq:firstorderform})
is.

%###\begin{figure}
%###\caption{Momentum constraint ${\cal C}_x$ versus time for evolutions
%###of a Painlev\'{e}-Gullstrand time slicing of a Schwarzschild black
%###hole using the Einstein-Christoffel system. Results are plotted for
%###several radial resolutions ranging from $N_r=10$ to $N_r=40$, fixed
%###angular resolution $\ell = 7$ and fixed time resolution $\Delta t =
%###0.015M$. Higher radial resolutions correspond to smaller errors.}
%###\label{fig:HW03-05-2001_ECMxl2}
%###\end{figure}
%###
%###\begin{figure}
%###\caption{Momentum constraint ${\cal C}_x$ versus time for 
%###the same evolutions shown in Figure~\ref{fig:HW03-05-2001_ECMxl2}
%###except $\eta=4/33$ and $\hat{z}=-1/4$, and we plot more radial resolutions.
%###If the outer boundary is moved out to $r=40M$, the run time extends to
%###$\sim 1300M$ for the same accuracy.}
%###\label{fig:LKWF09-19-2000A_ECMxl2}
%###\end{figure}

\end{document}